\documentclass[aps,floats]{revtex4}
\usepackage{amsmath,amssymb}
\usepackage{graphicx,epsfig}

\begin{document}
\bibliographystyle {plain}

\def\oppropto{\mathop{\propto}} 
\def\opsimeq{\mathop{\simeq}}
\def\opoverderline{\mathop{\overline}}
\def\operarrow{\mathop{\longrightarrow}}
\def\opsim{\mathop{\sim}}

\def\fig#1#2{\includegraphics[height=#1]{#2}}
\def\figx#1#2{\includegraphics[width=#1]{#2}}


\title{Many-body localization transition in a lattice model of interacting fermions : \\
statistics of renormalized hoppings in configuration space } 


 \author{ C\'ecile Monthus and Thomas Garel }
  \affiliation{ Institut de Physique Th\'{e}orique, CNRS and CEA Saclay,
 91191 Gif-sur-Yvette, France}

\begin{abstract}
We consider the one-dimensional lattice model of interacting fermions with disorder studied previously by Oganesyan and Huse [Phys. Rev. B 75, 155111 (2007)]. To characterize a possible many-body localization transition as a function of the disorder strength $W$, we use an exact renormalization procedure in configuration space that generalizes the Aoki real-space RG procedure for Anderson localization one-particle models [H. Aoki, J. Phys. C13, 3369 (1980)]. We focus on the statistical properties of the renormalized hopping $V_L$ between two configurations separated by a distance $L$ in configuration space (distance being defined as the minimal number of elementary moves to go from one configuration to the other). Our numerical results point towards the existence of a many-body localization transition at a finite disorder strength $W_c$. In the localized phase $W>W_c$, the typical renormalized hopping $V_L^{typ} \equiv e^{\overline{ \ln V_L }}$ decays exponentially in $L$ as $ (\ln V_L^{typ}) \simeq - \frac{L}{\xi_{loc}}$ and the localization length diverges as $\xi_{loc}(W) \sim (W-W_c)^{-\nu_{loc}}$ with a critical exponent of order $\nu_{loc} \simeq 0.45$. In the delocalized phase  $W<W_c$, the renormalized hopping remains a finite random variable as $L \to \infty$, and the typical asymptotic value $V_{\infty}^{typ} \equiv e^{\overline{ \ln V_{\infty} }}$ presents an essential singularity $(\ln V_{\infty}^{typ}) \sim - (W_c-W)^{-\kappa}$ with an exponent of order $\kappa \sim 1.4$. Finally, we show that this analysis in configuration space is compatible with 
the localization properties of
the simplest two-point correlation function in real space.

\end{abstract}

\maketitle

 \section{ Introduction }

Whereas Anderson localization phenomena \cite{anderson} are rather well
understood for a single particle
 (see the reviews \cite{janssenrevue,markos,mirlinrevue}),
the case of interacting particles in a random potential
 has remained much more challenging (see the review \cite{belitz}).
 In the field of disordered fermions, 
there has been for instance a lot of works 
 on  quantum Coulomb glasses (see for instance \cite{efros,ioffe,goethe}
and references therein), including
the debate on the existence of a metal-insulator transition
for interacting electrons in two dimensions (see \cite{vojta,berkovits2d,fleury}
and references therein). In the field of disordered bosons, many studies have been devoted
to the existence and properties of the superfluid-insulator transition  (see \cite{giamarchi,mfisher,refael,nattermann,ioffe_mezard,aleiner}
and references therein).
Recently, the idea to reformulate the many-body localization problem
as an Anderson localization problem in Fock space or in Hilbert space
has been very useful
\cite{levitov,berkovits98,leyronas,silvestrov,flambaum,berkovits03,gornyi,basko}.
In particular, this type of analysis has led to the prediction
that the conductivity of interacting electrons models
could exactly vanish in some finite region of parameters
in the absence of any external continuous bath \cite{gornyi,basko}.
The reason is that conduction mechanisms based on variable-range hopping
need a continuous bath to locally supply or absorb energy to permit hopping
between levels which are not exactly degenerate. 
Since quantum levels are discrete,
the many-particle system can fail to be an effective heat-bath for itself. 
Following these ideas, Oganesyan and Huse have proposed
that this type of many-body localization transition could be realized in
some one-dimensional lattice models of interacting fermions \cite{huse}.
 Unfortunately, the numerical study concerning 
the spectral statistics alone presented in Ref. \cite{huse}
has turned out to be not completely conclusive as a result of very
strong finite-size effects. Moreover, the analogy with Anderson localization
on high dimensional and Cayley tree indicates that 
criteria based on the level repulsion may not be very appropriate in this case
\cite{huse}. In the present paper, 
we propose to study the existence of a many-body localization
transition in the very same model of Ref. \cite{huse} 
by studying another type of observable which contains some information 
on the localization of eigenstates.
More precisely, we use an exact renormalization
 procedure in configuration space
to compute numerically the renormalized hopping
 between two configurations as a function of
their distance in configuration space. We present numerical results on 
the statistical properties of this renormalized hopping that point towards
the existence of a many-body localization transition at a finite disorder strength. 
In the localized phase, we measure a power-law divergence
of the localization length. In the delocalized phase, we measure an essential singularity
for the asymptotic renormalized hopping.
These scaling laws are thus reminiscent of the Anderson localization transition on the Cayley tree, which is the simplest example of Anderson transition in a space of infinite dimension, but the values of exponents are different.

The paper is organized as follows.
In Section \ref{secrg}, 
we describe how many-body localization models can be studied numerically
via an exact renormalization procedure ('RG') in configuration space
that generalizes Aoki real-space RG procedure for Anderson localization one-particle models. In Section \ref{secnume}, we present our numerical results 
 for a one-dimensional lattice model of interacting fermions.
In section \ref{secdiscussion}, we discuss
 the similarities and differences with the scaling laws
of Anderson localization on the Cayley tree, 
and propose a specific form of finite-size scaling analysis.
In section \ref{sec-corre}, we present independent numerical results concerning 
the simplest real-space two-point correlation function, to test
 the compatibility with our results obtained in configuration space.
Our conclusions are summarized in section \ref{secconclusion}.

\section{ Exact renormalization procedure in configuration space }

\label{secrg}

In this section, we briefly summarize the 
Aoki real space renormalization ('RG') method for one-particle localization models
before we describe its generalization for many-body problems.

\subsection{ Reminder on Aoki real space RG for one-particle localization models }

For Anderson localization models, there exists an exact real-space renormalization procedure at fixed energy $E$ which preserves the Green functions of the remaining sites
\cite{aoki80,aoki82,aokibook,lambert,
us_twopoints,us_rgloc}.

The renormalization (RG) procedure 
can be applied to any Anderson localization model of the generic form
\begin{eqnarray}
H =   \sum_{i,j}  V_{i,j} \vert i > < j \vert
\label{Hgene}
\end{eqnarray}
 where $V_{i,i}$ is the on-site energy of site $i$,
and where $V_{i,j}$ is the hopping between the sites $i$ and $j$.
Upon the elimination of site $i_0$ in the
Schr\"odinger equation at energy $E$
\begin{eqnarray}
E \psi(i_0) = V_{i_0,i_0} \psi(i_0) +\sum_{j \ne i_0} V_{i_0,j} \psi(j)
\label{psii0}
\end{eqnarray}
the remaining sites satisfy the
Schr\"odinger equation at energy $E$ with the renormalized parameters  
\begin{eqnarray}
V_{i,j}^{new} = V_{i,j} + \frac{V_{i,i_0} V_{i_0,j} }{E-V_{i_0,i_0}}
\label{rulev}
\end{eqnarray}
These renormalization equations are exact since they are based on
elimination of the variable $\psi(i_0)$ in the Schr\"odinger Equation.
As stressed by Aoki \cite{aoki80,aoki82}, the RG rules preserve the Green function
for the remaining sites.
This means for instance that if external leads
are attached to all surviving sites, the scattering properties
will be exactly determined using the renormalized parameters 
(see \cite{us_rgloc} for more details).
In particular, the renormalized hopping between the last two surviving sites
(after all other sites have been decimated) determines the two-point Landauer
transmission between leads attached to these two points \cite{us_rgloc} :
it decays exponentially with the distance in the localized phase,
it remains finite in the delocalized phase, 
and at the critical point it becomes multifractal.

\subsection{Generalization in configuration space for many-body localization models }

The above RG procedure has the following natural generalization
for many-body models.
Let us denote ${\cal C}$ 
a configuration of the many-body problem
to write the Hamiltonian as
\begin{eqnarray}
H_{many} = \sum_{{\cal C}_i, {\cal C}_j}
  V_{{\cal C}_i,{\cal C}_j} \vert {\cal C}_i > < {\cal C}_j \vert
\label{Hmany}
\end{eqnarray}
Then the Schr\"odinger equation projected onto the configuration 
${\cal C}_{i_0}$
\begin{eqnarray}
E \psi({\cal C}_{i_0}) = \sum_{{\cal C}_j} V_{{\cal C}_{i_0},{\cal C}_j}
 \psi({\cal C}_j)
\label{psii0config}
\end{eqnarray}
allows to eliminate $ \psi({\cal C}_{i_0})$. The remaining configurations satisfy the Schr\"odinger equation with renormalized parameters
\begin{eqnarray}
V_{{\cal C}_i,{\cal C}_j}^{new} = V_{{\cal C}_i,{\cal C}_j}
 + \frac{ V_{{\cal C}_i,{\cal C}_{i_0}} 
 V_{{\cal C}_{i_0},{\cal C}_{j}}  }
{E- V_{{\cal C}_{i_0},{\cal C}_{i_0}} }
\label{rulevconfig}
\end{eqnarray}
These rules in configuration space have been
already used for the two-particle 1D Anderson tight-binding 
model \cite{leadbeater}. In the following, we apply them to
a model of interacting fermions that we now describe.

\subsection{ Application to the interacting fermions model of Ref. \cite{huse} }

\label{sec-model}

In numerical studies of
 quantum problems containing both interactions and disorder,
 it is natural to consider first the spatial dimension $d=1$.
The simplest model is then a chain of spinless fermions with nearest-neighbor
interaction and on-site disorder
 (see for instance \cite{schmitteckert,molina,MacK} and references therein).
This type of model can also be studied in the language of quantum spin chains
as in Ref. \cite{znidaric}, where a powerful time-dependent Density Matrix Renormalization Group method has been used to characterize the many-body localized phase.
In this paper, we consider the same class of model, but with 
 second-neighbor hopping in addition, as in Ref. \cite{huse}.
More precisely, the model of Ref. \cite{huse}
is defined by the following Hamiltonian on a one-dimensional lattice
of $L$ sites with periodic boundary conditions
\begin{eqnarray}
H= \sum_{i=1}^L \left[ w_i n_i + V \left( n_i - \frac{1}{2}\right) 
\left( n_{i+1}- \frac{1}{2}\right) + c_i^{\dagger} c_{i+1}
 + c_{i+1}^{\dagger} c_{i} + c_i^{\dagger} c_{i+2} + c_{i+2}^{\dagger} c_{i}
\right]
\label{hamilton}
\end{eqnarray}
with the usual notations :

(i) $n_i=c_i^{\dagger} c_i$ represents
 the number of spinless fermion on site $i$
and can take only two values
 ($0$ if the site is empty or $1$ if the site is occupied).
The many-body Hilbert space has thus for dimension 
\begin{eqnarray}
 {\cal N}_L= 2^L
\label{dimconfig}
\end{eqnarray}
The spinless character has been chosen to reach bigger sizes $L$
for a given value of the Hilbert space dimension ${\cal N}_L$ \cite{huse}.

(ii) the on-site energies $w_i$ are independent Gaussian variables with zero mean and variance $W^2$, i.e. $W$ measures the disorder strength (we have not used
the 'microcanonical constraint' of Ref. \cite{huse} consisting in the requirement
that $(1/L) \sum w_i^2$ should be exactly $W^2$ to reduce statistical uncertainties).

(iii) the nearest-neighbor interaction is chosen to be $V=2$,
 the hopping terms between nearest-neighbors and second-neighbors
are chosen to be $t=t'=1$. The second-neighbor hopping is included 
to have non-integrability at zero randomness, 
see \cite{husepur} for more details
on the properties of the model in the  zero-disorder limit.

(iv) the total number of particles is conserved :
we study the case of half-filling with $L/2$ particles for $L$ sites
as in \cite{huse}. The dimension of the Hilbert space is then given by the 
binomial coefficient
\begin{eqnarray}
 {\cal N}_L^{half-filling}= \binom{L}{ \frac{L}{2}} \oppropto_{L \to +\infty} \frac{2^L}{\sqrt L}
\label{binomial}
\end{eqnarray}
Physically, the important point is that at leading order,
it still grows exponentially in $L$.

In summary, we consider in this paper the model of Ref. \cite{huse}
with exactly the same values of parameters, but we study
another observable to detect the possible many-body localization transition.
We have applied the RG procedure in configuration space described above,
to obtain, in each disordered sample of even size $L$, the renormalized hopping
$V_L$ at some energy $E$ 
between two configurations ${\cal C}_A,$ and ${\cal C}_B$ after the decimation of all other configurations
\begin{eqnarray}
V_L \equiv V_{{\cal C}_A,{\cal C}_B} (E)
\label{defvL}
\end{eqnarray}
We have made the following choices :

(c1) we consider the zero-energy case $E=0$, because it represents
the center of the many-body energy levels. Indeed, in Anderson localization models, it is well known that energy levels 
near the center are the more favourable
to delocalization : if these states are localized, one expects that all other states will also be localized. 

(c2) we have chosen the following configurations :
the configuration ${\cal C}_A$ has
 all even sites occupied and all odd sites empty,
whereas 
the configuration ${\cal C}_B$ has
 all odd sites occupied and all even sites empty.
Their distance in configuration space is thus $L/2$ (the minimal path to
 go from configuration ${\cal C}_A$ to configuration ${\cal C}_B$ requires
$L/2$ elementary moves).
In the absence of disorder, the model is known to be conducting
(see \cite{husepur} for a detailed study of conductivity properties) ;
the two configurations ${\cal C}_A$ and  ${\cal C}_B$ 
are equivalent up to a translation of one lattice site, 
and are thus expected to be connected by a finite renormalized hopping.
 In the presence of disorder, these two configurations are 
not equivalent anymore, and one expects that the renormalized hopping $V_L$ 
will become exponentially small in $L$ for sufficiently strong disorder.

This choice (c2) of alternate configurations ${\cal C}_A$ and  ${\cal C}_B$
can be questioned in various ways. For instance, if one wishes to maximize
the distance in Fock space, one obtains the configurations where all
particles are on the first half or on the second half : 
physically, it is however clear that these two configurations
 are not typical because they are extremely inhomogeneous and because
only 4 particles (2 particles at each boundary of the macroscopic cluster)
can move for arbitrary large $L$ (instead of an extensive number of particles for typical configurations).
More generally, in contrast to usual Anderson localization models
where all sites are equivalent, a new difficulty that arises in 
many-body localization models
is that all configurations are not equivalent : 
configurations have different hopping connectivities, 
and different interaction energies, so that the configuration space has
already an inhomogeneous structure even before
the introduction of disorder variables.
Since an extensive study of the renormalized hoppings in this complicated 
configuration space is not really possible, we have decided to consider
only the choice (c2) of alternate configurations ${\cal C}_A$ and  ${\cal C}_B$
in the remaining of this paper. However, to show that our results are meaningful
and do not really depend of our precise choice (c2), we present in
section \ref{sec-corre} independent numerical results concerning 
the simplest real-space two-point correlation function.

\section{ Numerical results for the  
interacting fermions model of Ref. \cite{huse} }

\label{secnume}

In this section, we describe our numerical results concerning the statistics of $V_L$ at $E=0$ between the alternate configurations ${\cal C}_A$
and ${\cal C}_B$ described above for even sizes $4 \leq L \leq 12$
with corresponding statistics $15.10^7 \leq n_s(L) \leq 3650$ 
of disordered samples
(we have also data corresponding to $L=14$ with $n_s=100$ samples, but this statistics has turned out to be insufficient for most purposes).

\subsection{ Analysis of the localized phase }

\begin{figure}[htbp]
 \includegraphics[height=6cm]{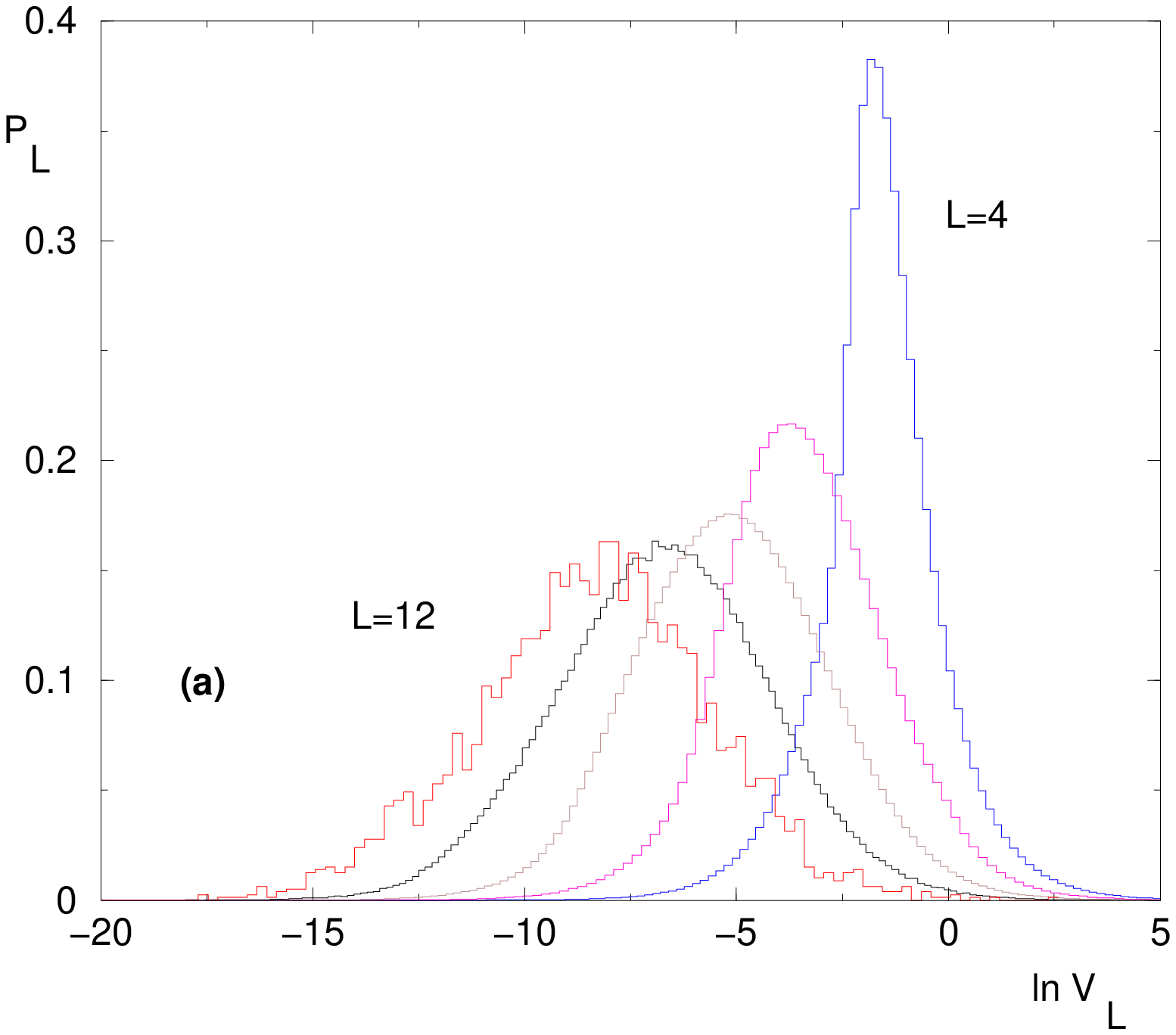}
\hspace{1cm}
 \includegraphics[height=6cm]{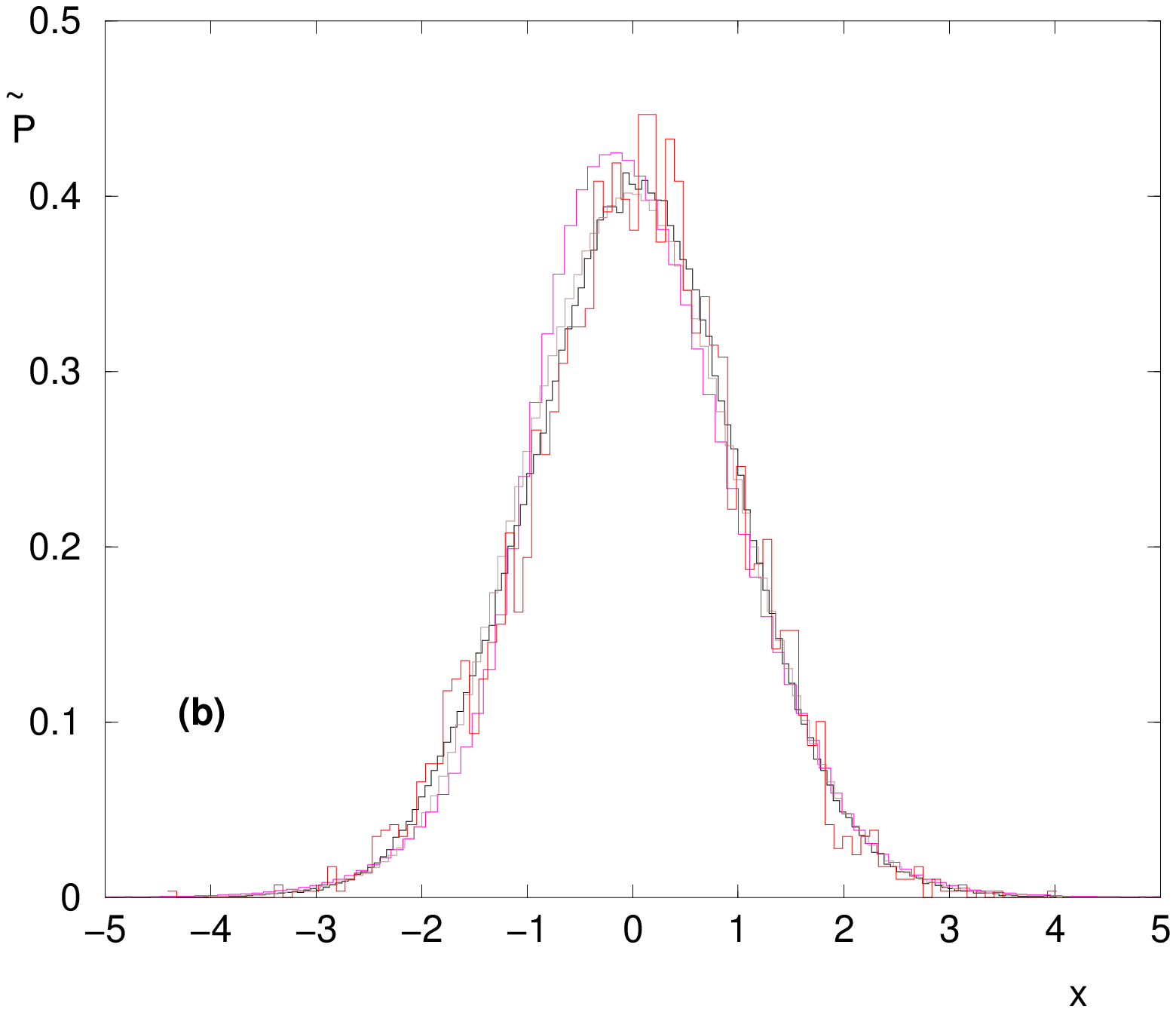}
\caption{ (Color on line) Statistics of the renormalized hopping $V_L$ in the localized phase 
(data for the disorder strength $W=20$) :
(a) Probability distribution $P_L(\ln V_L)$
of the logarithm of renormalized hopping $V_L$ for the sizes $L=4,6,8,10,12$.
(b) same data for the rescaled variable $x$ of Eq. \ref{defxrescal}
for the sizes $L=6,8,10,12$ : 
the convergence towards a fixed rescaled distribution ${\tilde P}(x) $ is rapid
(we have only excluded the smallest size $L=4$ that was too different). 
  }
\label{fighistoloc}
\end{figure}

For strong disorder, we find that the renormalized hopping $V_L$ introduced in Eq. \ref{defvL}
flows towards smaller and smaller values as $L$ increases. 
As an example for $W=20$, we show on Fig. \ref{fighistoloc} the probability distributions
$P_L(\ln V_L)$  of the variable $\ln V_L$ over the disordered samples of a given size
 $L=4,6,8,10,12$ : the regular shift of these histograms towards smaller values is clear.
 We show on Fig. \ref{fighistoloc} (b) the same data for the rescaled variable 
 \begin{eqnarray}
x \equiv \frac{ \ln V_L - \overline{ \ln V_L} }{ \Delta_L}
\label{defxrescal}
\end{eqnarray}
where $\overline{ \ln V_L}$ is the averaged value and where $\Delta_L$ is the width of the distribution
$P_L(\ln V_L)$. One can see on
Fig. \ref{fighistoloc} (b) that the histograms of the
rescaled variable $x$ coincide within statistical fluctuations for $L=6,8,10,12$
(we have only excluded the smallest size $L=4$ that was too different) :
this shows that the convergence towards a stable rescaled distribution ${\tilde P}(x) $
is rapid for this model.

\begin{figure}[htbp]
 \includegraphics[height=6cm]{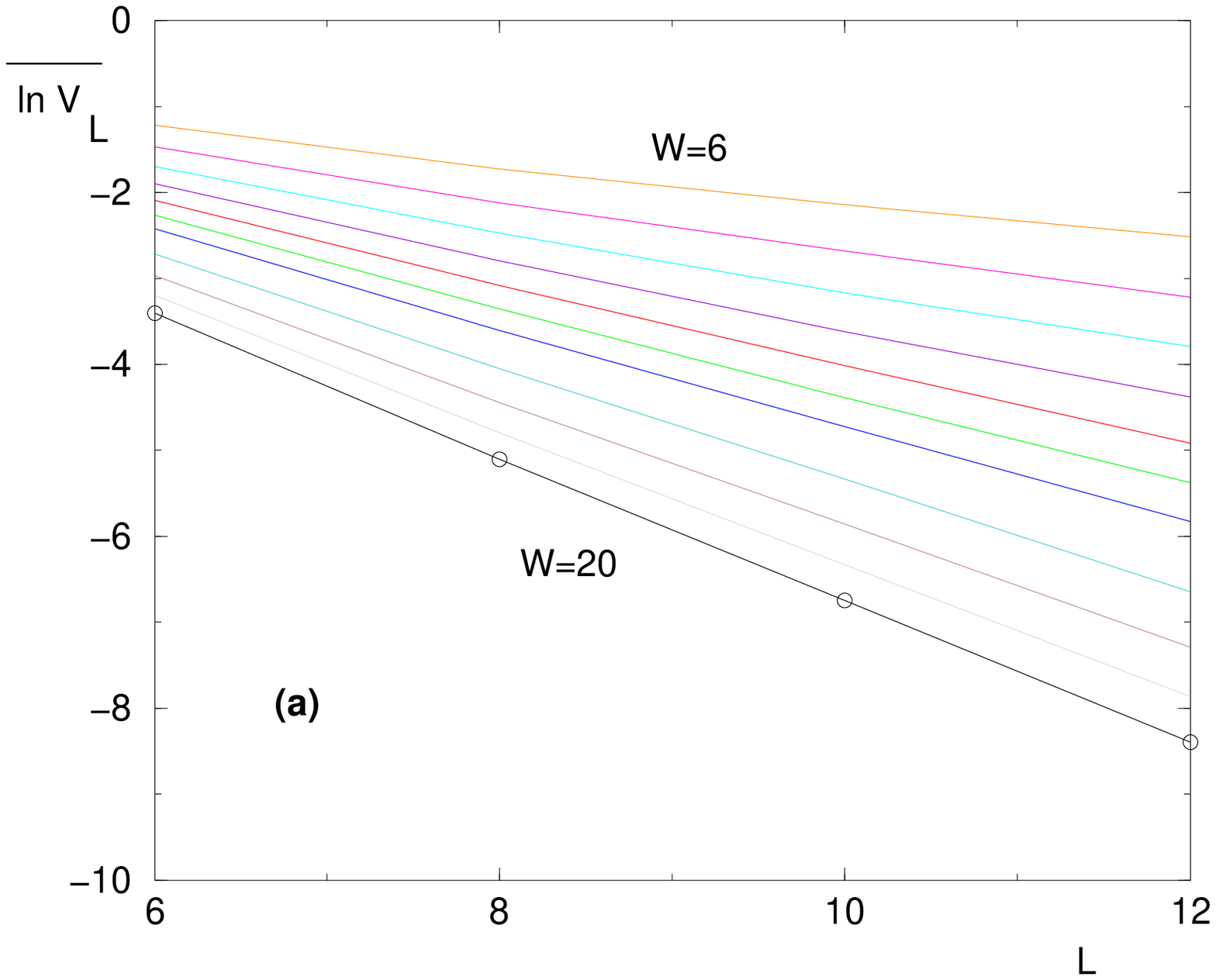}
\hspace{2cm}
\includegraphics[height=6cm]{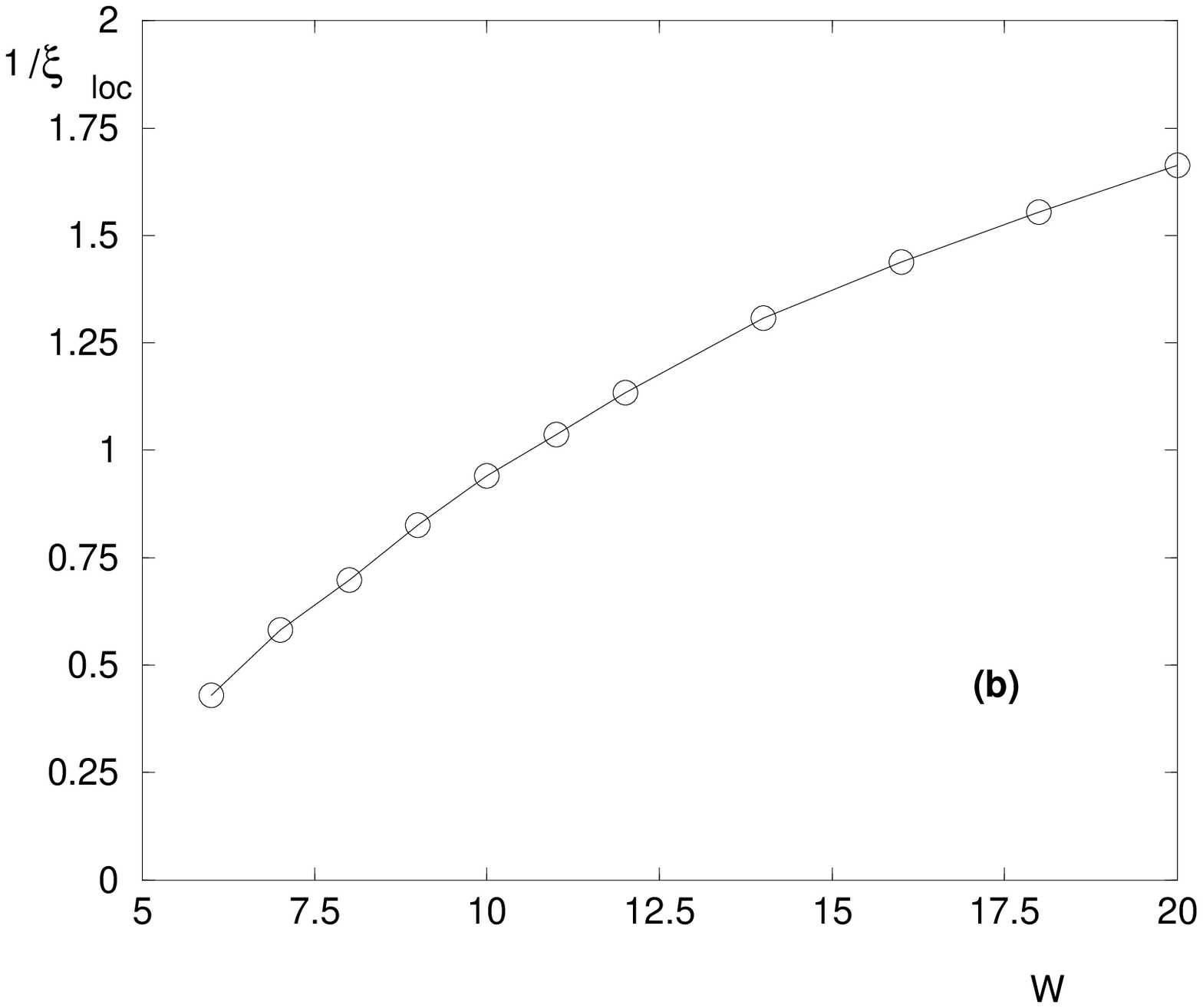}
\caption{ (Color on line) Exponential decay of the typical renormalized hopping 
$V_L^{typ} \equiv e^{\overline{ \ln V_L }}$ in the localized phase  :
(a) Linear decay of $\overline{ \ln V_L }$ 
as a function of $L$  (see Eq. \ref{transloc}).
(b) Behavior of the slope $1/\xi_{loc}(W)$ (inverse of the localization length $\xi_{loc}(W)$ of Eq. \ref{transloc}) as a function
 of the disorder strength $W$  }
\label{figtransloc}
\end{figure}

We show on Fig. \ref{figtransloc} (a) the decay with $L$ 
of the disorder-average $\overline{ \ln V_L}$
for various disorder strengths in the range $6 \leq W \leq 20$ : these curves correspond
to an exponential decay of the typical value 
$V_L^{typ} \equiv e^{\overline{ \ln V_L }}$
with respect to the distance $(L/2)$ in configuration space
\begin{eqnarray}
\ln (V_L^{typ}) \equiv \overline{ \ln V_L(W>W_c) }\
 \opsimeq_{L \to \infty} - \frac{(L/2)}{\xi_{loc}(W)}
\label{transloc}
\end{eqnarray}
where $\xi_{loc}(W)$ represents the localization length that diverges
at the delocalization transition
\begin{eqnarray}
\xi_{loc}(W) \opsimeq_{W \to W_c^+} (cst) (W-W_c)^{-\nu_{loc}}
\label{xiloc}
\end{eqnarray}
We show on Fig. \ref{figtransloc} (b) our numerical result
for the slope  $1/\xi_{loc}(W)$ as a function of the disorder strength $W$
in the region $6 \leq W \leq 20$ (below $W=6$ we cannot estimate the linear slope anymore).
A three-parameter fit of the form of Eq. \ref{xiloc} yields 
a critical point in the range
\begin{eqnarray}
5.2  \leq W_c \leq 5.9
\label{critiloc}
\end{eqnarray}
and a critical exponent around
\begin{eqnarray}
\nu_{loc} \simeq 0.45
\label{nuloc}
\end{eqnarray}

\subsection{ Analysis of the delocalized phase }

\label{secdeloc}

\begin{figure}[htbp]
 \includegraphics[height=6cm]{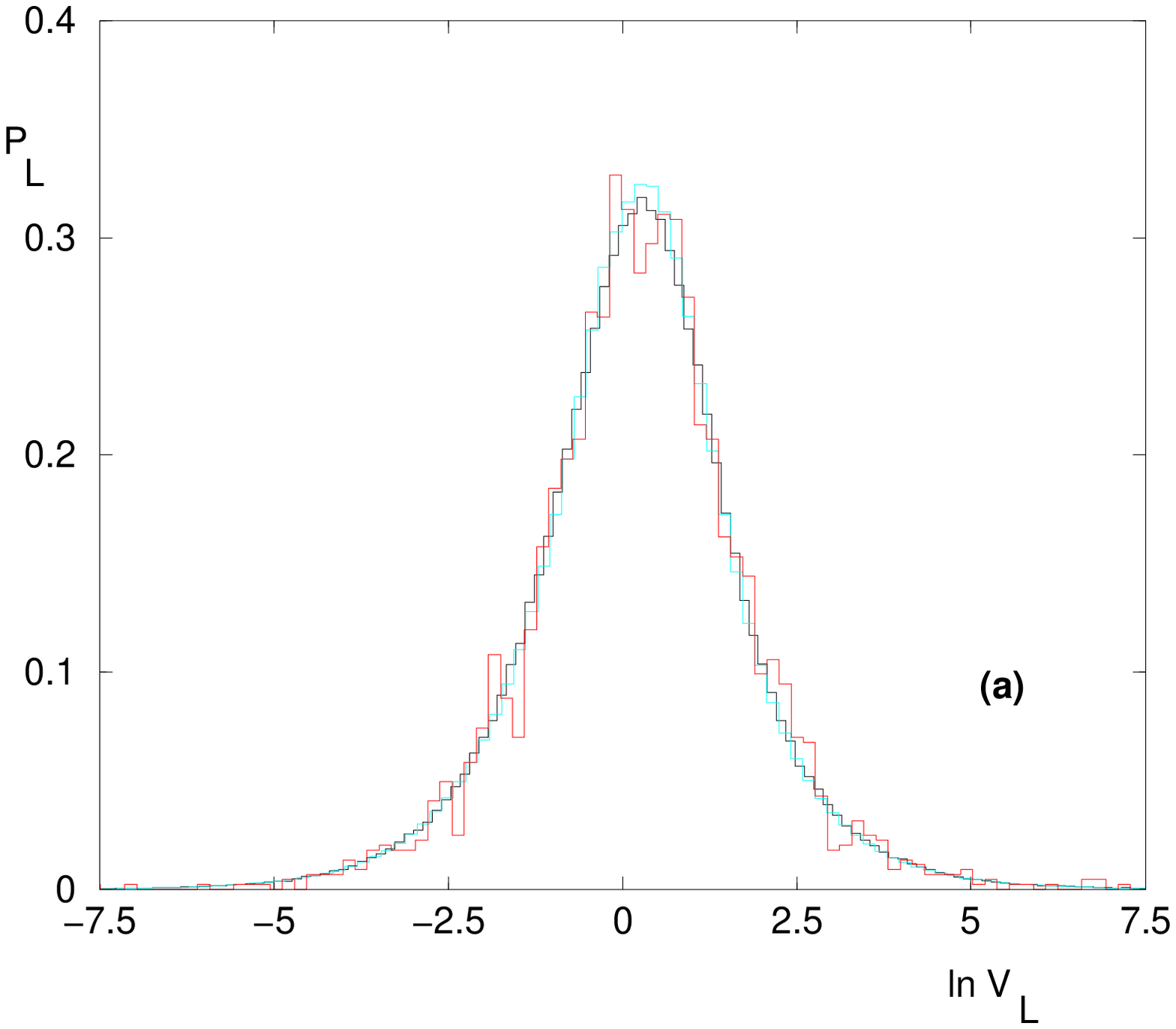}
\hspace{1cm}
 \includegraphics[height=6cm]{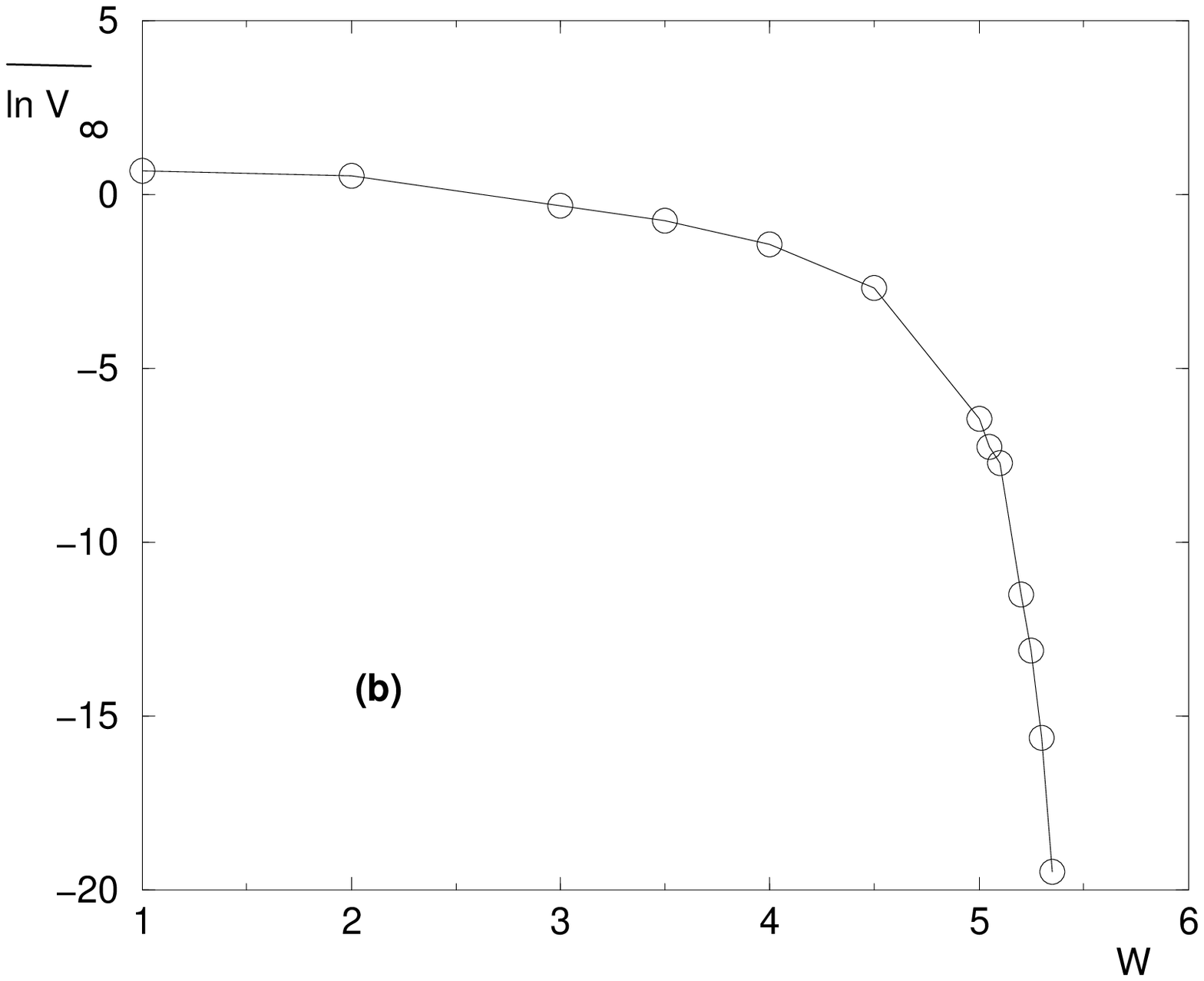}
\caption{ (Color on line) Statistics of the renormalized hopping $V_L$ 
in the delocalized phase 
(a) For the disorder strength $W=2$, the probability distribution $P_L(\ln V_L)$
of the logarithm of renormalized hopping $V_L$ remains the same for the sizes $L=8,10,12$ 
(we have excluded the smallest sizes $L=4,6$ that were a bit different). 
This should be compared with  Fig. \ref{fighistoloc} (a) corresponding to the localized phase.
(b) Behavior of the typical asymptotic renormalized hopping $V_{\infty}^{typ} \equiv e^{\overline{ \ln V_{\infty} }}$ : $\ln V_{\infty}^{typ} = \overline{\ln V_{\infty}} $ as a function of the disorder strength $W$. 
  }
\label{figdeloc}
\end{figure}

For weak disorder, we find that the renormalized hopping $V_L$ introduced in Eq. \ref{defvL}
remains a finite random variable $V_{\infty}$ finite as $L$ increases. 
As an example for $W=2$, we show on Fig. \ref{figdeloc} the probability distributions
$P_L(\ln V_L)$  of the variable $\ln V_L$ over the disordered samples of a given size
 $L=8,10,12$ (for clarity we have excluded the smallest sizes
 $L=4,6$ that were too different) : 
 it is clear that these histograms coincide up to statistical fluctuations.
This should be compared with  Fig. \ref{fighistoloc} (a) corresponding to the localized phase for $W=20$.
In the delocalized phase, the typical renormalized hopping 
$V_{\infty}^{typ} \equiv e^{\overline{ \ln V_{\infty} }}$ thus remains finite
\begin{eqnarray}
\overline{ \ln V_L(W<W_c,L) } \opsimeq_{L \to \infty} 
 \overline{ \ln V_{\infty}(W<W_c) } \ \ \ {\rm finite }
\label{transdeloc}
\end{eqnarray}
We show on Fig. \ref{figdeloc} (b) our numerical estimates of
the asymptotic value $\overline{ \ln V_{\infty}(W<W_c) }$ as a function of $W$.
We find that our data are compatible with an 
 essential singularity behavior of the typical asymptotic hopping $V_{\infty}^{typ}$
\begin{eqnarray}
\ln V_{\infty}^{typ} (W<W_c)  \equiv \overline{ \ln V_{\infty}(W<W_c) }
 \opsimeq_{W \to W_c^-} -  (cst)(W_c-W)^{- \kappa}  
 \label{essentialsingularitytransdeloc}
\end{eqnarray}
A three-parameter fit of this form  yields 
a critical point in the range 
\begin{eqnarray}
5.5 \leq W_c \leq 5.7
\label{critideloc}
\end{eqnarray}
and an essential singularity exponent around
\begin{eqnarray}
 \kappa \simeq 1.4
\label{kappa}
\end{eqnarray}

Essential singularities in transport 
properties have already been found in various
disordered models, in particular in Anderson localization on the Cayley tree
(see \cite{MirlinBethe,us_cayley} and references therein)
and in superfluid-insulator transitions of disordered bosons 
(see \cite{giamarchi,mfisher,refael,ioffe_mezard} and references therein).

\subsection{ Conclusion of the numerical study }

\begin{figure}[htbp]
 \includegraphics[height=6cm]{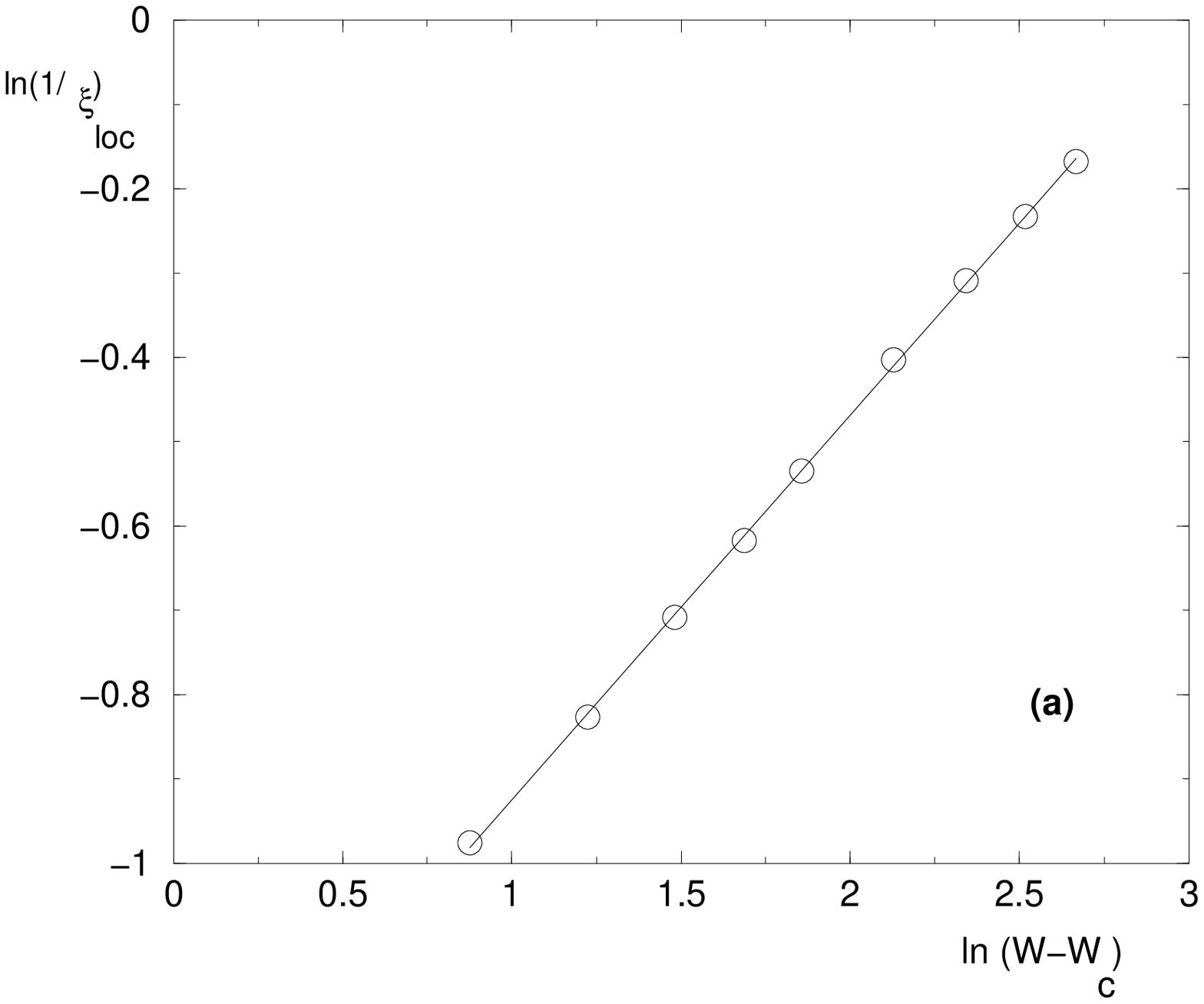}
\hspace{1cm}
 \includegraphics[height=6cm]{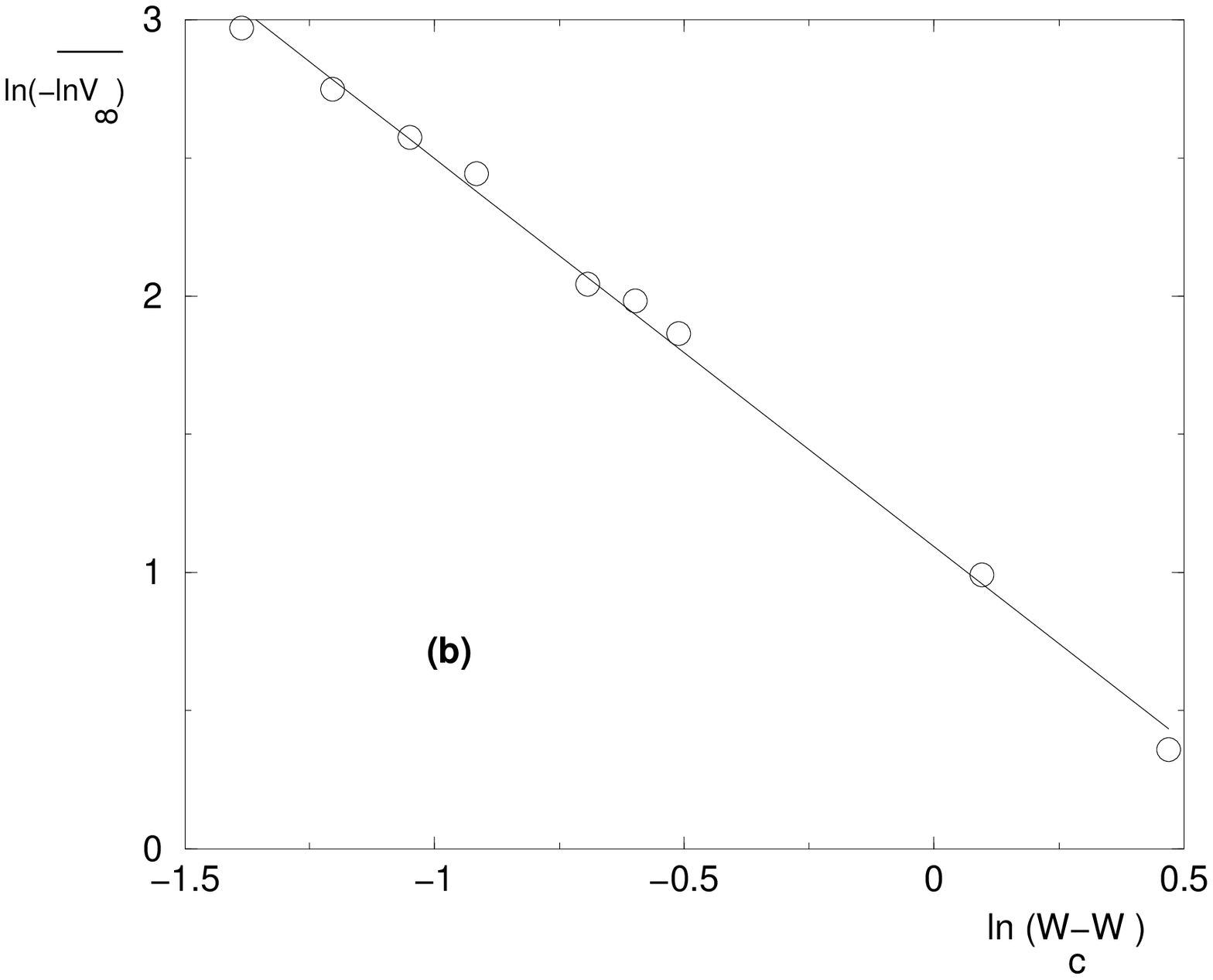}
\caption{ Critical behaviors obtained for a critical point at the value $W_c=5.6$ 
(a) Localized phase $W>W_c$ : the plot of $\ln (1/\xi_{loc} (W))$ as a function of 
$\ln (W-W_c)$ corresponds to the slope  $\nu_{loc} \simeq 0.45$
(see Eq. \ref{xiloc})
(b) Delocalized phase $W<W_c$ : the plot of $\ln (- \overline{ \ln V_{\infty}})$
as a function of $\ln (W_c-W)$ corresponds to the slope 
 $\kappa \simeq 1.4$ (see Eq. \ref{essentialsingularitytransdeloc}) 
  }
\label{figcriti}
\end{figure}

In summary, our numerical data are compatible
 with a many-body localization transition for the model of Eq. \ref{hamilton}. 
The global best value for the critical point seems to be
\begin{eqnarray}
 W_c \simeq 5.6
\label{wc}
\end{eqnarray}
which is somewhat smaller than the critical value suggested by 
the level statistics study of Ref. \cite{huse}.
A possible reason for this slight difference could be that the 
level statistics study of Ref. \cite{huse} is based on all levels of all energies,
that could mix contributions of various types of states (delocalized, localized, and critical), whereas we have chosen to work at the fixed energy $E=0$ (center of the many-body energy levels). Anyway, taking into account the large 
uncertainties on $W_c$ as estimated from small system sizes, we feel that
the two studies point towards the same region of disorder strength $W$.

For the value of Eq. \ref{wc}, 
we show the log-log plots of the critical behaviors
on Fig. \ref{figcriti}.
On Fig. \ref{figcriti} (a), we show the divergence of the localization length $\xi_{loc}(W)$ in the localized phase $W>W_c$ : 
the slope  $\nu_{loc} \simeq 0.45$ (see Eq. \ref{xiloc}).
On Fig. \ref{figcriti} (b), we show the essential singularity
of the typical asymptotic hopping in the delocalized phase $W<W_c$ :
the slope corresponds to the exponent $\kappa \simeq 1.4$
(see Eq. \ref{essentialsingularitytransdeloc}).
Of course, these values are not expected to be precise, since they have been
obtained from small system sizes and some fitting/extrapolation procedures
from the raw data. Nevertheless, the emergence of reasonable scaling laws
is encouraging. In the following section, we discuss
the similarity with the scaling laws 
 that appear for Anderson localization on the Cayley tree.

\section{ Discussion : similarities and differences with Anderson localization on the Cayley tree}

\label{secdiscussion}

\subsection{ Analogy with Anderson localization on the Cayley tree}

As recalled in the introduction, the reformulation of the many-body localization problem
as an Anderson localization problem in Fock space or in Hilbert space
has been very useful
\cite{levitov,berkovits98,leyronas,silvestrov,flambaum,berkovits03,gornyi,basko}.
The idea is to analyse whether there exists an Anderson localization in configuration space, and to study the consequences for real-space properties.
The geometry of configuration space is usually very different from 
the regular finite-dimensional lattices
 considered in Anderson one-particle localization models,
and has been argued to be qualitatively similar to the Cayley tree 
 \cite{levitov,berkovits98,silvestrov,gornyi}.
 Since  Anderson localization on the Cayley tree
has been studied for a long time as a mean-field limit
\cite{abou,Kun_Sou,MirlinBethe,efetovbook,DR,MD,us_cayley},
results and methods have been then
 borrowed to analyse many-body localization
properties of quantum dots  \cite{levitov,berkovits98,silvestrov,gornyi}.
 This approximation by a tree structure has been
however sometimes criticized \cite{leyronas}.
Indeed, in many-body localization models,
the Fock space or Hilbert space is never exactly a tree, 
and thus the approximation by a
 Cayley tree has been proposed as a simplifying approximation
to obtain an exactly solved model \cite{levitov} (note however that in \cite{gornyi}, it has been argued that an effective Cayley tree structure should actually well capture the properties of low-dimensional electronic models).
But independently of the technical convenience of the tree structure,
we believe that the physically important property in this analogy
is the 'infinite-dimension' property, defined as the exponential growth
of the configuration space ${\cal N_L}$ with the real-space linear size $L$
\begin{eqnarray}
 {\cal N}_L \propto e^{ (cst ) L}
\label{expo}
\end{eqnarray}
(whereas in finite dimension $d$, 
the configuration space of a single particle grows as
a power-law $ {\cal N}_L \propto L^d$). 
As argued in \cite{MFdc=infty}, 
it is the exponential growth of Eq. \ref{expo}
which is directly responsible for the presence of
 essential singularities of transport
properties, whereas finite-dimensional lattices 
are characterized by power-law singularities.
From the point of view of Anderson one-particle models, 
the Cayley tree is thus
rather 'pathological' since $d=\infty$ turns out to be a singular point,
and the upper critical dimension is considered to be 
$d_c=+\infty$ \cite{MFdc=infty}.
From the point of view of many-body localization however, 
the exponential growth of Eq.
\ref{expo} is the rule (see for instance Eq. \ref{dimconfig}), 
and thus the scaling behaviors that appear on the Caylee tree
are instructive, as an example of Anderson localization on a space of
infinite dimension. 
 In particular, this analysis suggests 
some specific form of finite-size scaling as we now recall.

\subsection{ Specific form of finite-size scaling in the critical region}

\begin{figure}[htbp]
 \includegraphics[height=6cm]{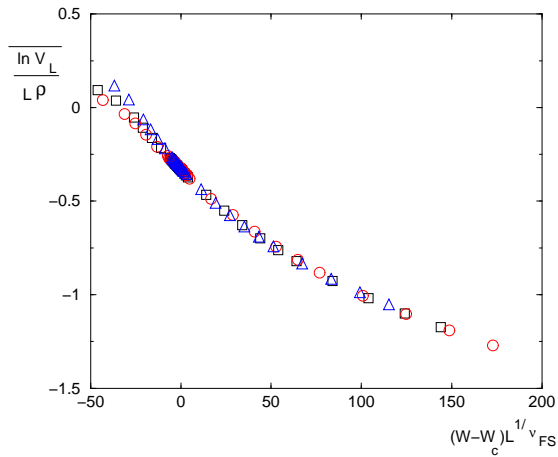}
\caption{ (Color on line) Finite-size scaling analysis of our numerical data
for the sizes $L=8$ (triangles), $L=10$ (squares) and $L=12$ (circles)
according to the form of Eq. \ref{transfss} with the values
$\rho = 0.76$ (see Eq. \ref{transrhonume}) and $\nu_{FS} = 1.85$
(see Eq. \ref{nuFStrans}) : this plot is rather 
convincing at criticality and in the localized phase $W>W_c$,
whereas stronger corrections to scaling seem to be present in the delocalized
phase $W<W_c$.
  }
\label{figfss}
\end{figure}

As in our recent study of the Landauer transmission for Anderson localization on the Cayley tree
\cite{us_cayley}, it is natural to assume 
some finite-size scaling in the critical region of the form
\begin{eqnarray}
\overline{ \ln V_L(W)} \opsimeq - L^{\rho} G \left( L^{1/\nu_{FS}} 
(W_c-W)  \right)
\label{transfss}
\end{eqnarray}
where the finite-size scaling exponent $\nu_{FS}$ is different from the
localization length exponent $\nu_{loc}$.
(This is in contrast with the scaling theory of localization in finite
dimension $d$, where the finite-size scaling is governed by $\nu_{loc}$).

The matching of Eq. \ref{transfss} with the 
localized phase  (see Eq. \ref{transloc} and Eq. \ref{xiloc}) yields
\begin{eqnarray}
 \nu_{loc}=  (1-\rho)\nu_{FS} 
\label{nutyprelation}
\end{eqnarray}
and the matching with the delocalized phase 
(Eq. \ref{essentialsingularitytransdeloc}) yields
\begin{eqnarray}
\kappa=  \rho \nu_{FS}
\label{kapparelation}
\end{eqnarray}

By consistence, the finite-size correlation length exponent $\nu_{FS}$ is then given by
\begin{eqnarray}
\nu_{FS} = \nu_{loc}+\kappa
\label{nuFStrans}
\end{eqnarray}
In an exactly solved travelling/non-travelling phase transition
 where the same type
of finite-size scaling occurs \cite{simon}, the physical interpretation of
 the finite-size scaling exponent 
$\nu_{FS}$ is that it governs the relaxation rate towards the finite value
in the non-travelling phase. For Anderson localization on the Cayley tree,
we have checked that this interpretation holds \cite{us_cayley}.
For the present many-body localization transition, 
this property cannot be checked with our numerical data limited to small sizes.  

Exactly at criticality, we thus expect the following
stretched exponential decay of the typical renormalized hopping
\begin{eqnarray}
\overline{\ln V_{L} (W_c) } \simeq -  L^{\rho}
\label{transcriti}
\end{eqnarray}
where the exponent $\rho$ is related to the other exponents by
(see the scaling relations of Eqs \ref{nutyprelation} and \ref{kapparelation} )
\begin{eqnarray}
\rho = \frac{ \kappa}{\nu_{FS}} = \frac{ \kappa}{\kappa + \nu_{loc}}
\label{transrho}
\end{eqnarray}
From our previous estimates of the exponents  $\nu_{loc} \simeq 0.45$ 
(Eq. \ref{nuloc}) and $\kappa \simeq 1.4$ (Eq. \ref{kappa}), 
this would correspond to a numerical value of order
\begin{eqnarray}
\rho  \simeq 0.76
\label{transrhonume}
\end{eqnarray}

We show on Fig. \ref{figfss}
the finite-size scaling analysis of our numerical data
according to the form of Eq. \ref{transfss} with the values
$\rho = 0.76$ and $\nu_{FS} = 1.85$ obtained by consistency from our previous
estimates of $\nu_{loc}$ and $\kappa$ : the data collapse 
seems satisfactory at criticality and in the localized phase $W>W_c$,
whereas stronger corrections to scaling seem to be present in the delocalized
phase $W<W_c$.

In summary of this discussion, we propose that the scaling laws of many-body localization transitions are generically similar to the scaling laws observed for Anderson localization on the Cayley tree, as a consequence of the 
infinite-dimension property of Eq. \ref{expo}. However, besides this qualitative analogy, one should not expect an exact equivalence in general,
and in particular the critical exponents $(\nu_{loc},\nu_{FS},\kappa,\rho)$
are not expected to be the same as those of the Cayley tree.

\section{ Numerical results concerning the simplest real-space two-point correlation function }

\label{sec-corre}

\begin{figure}[htbp]
 \includegraphics[height=6cm]{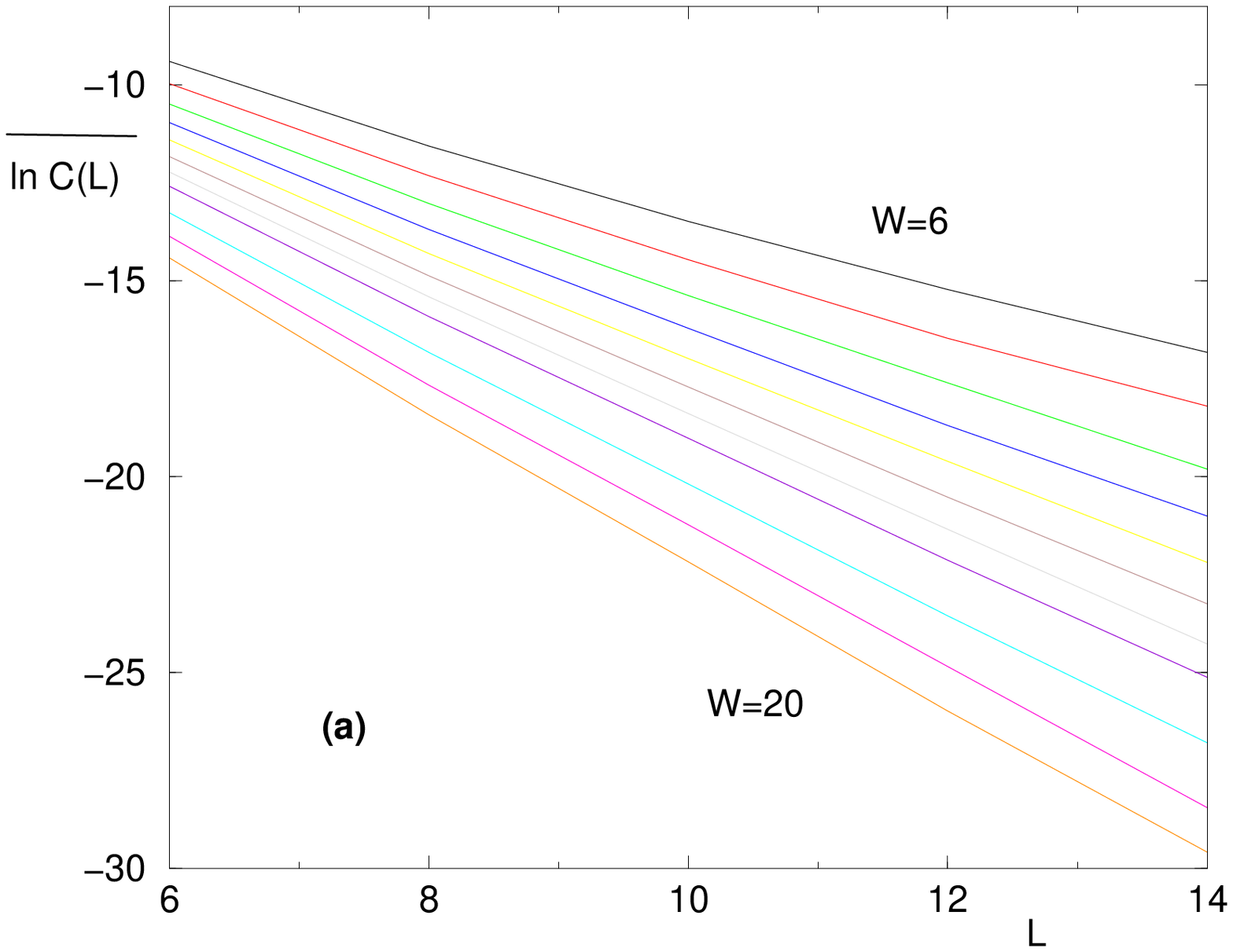}
\hspace{1cm}
 \includegraphics[height=6cm]{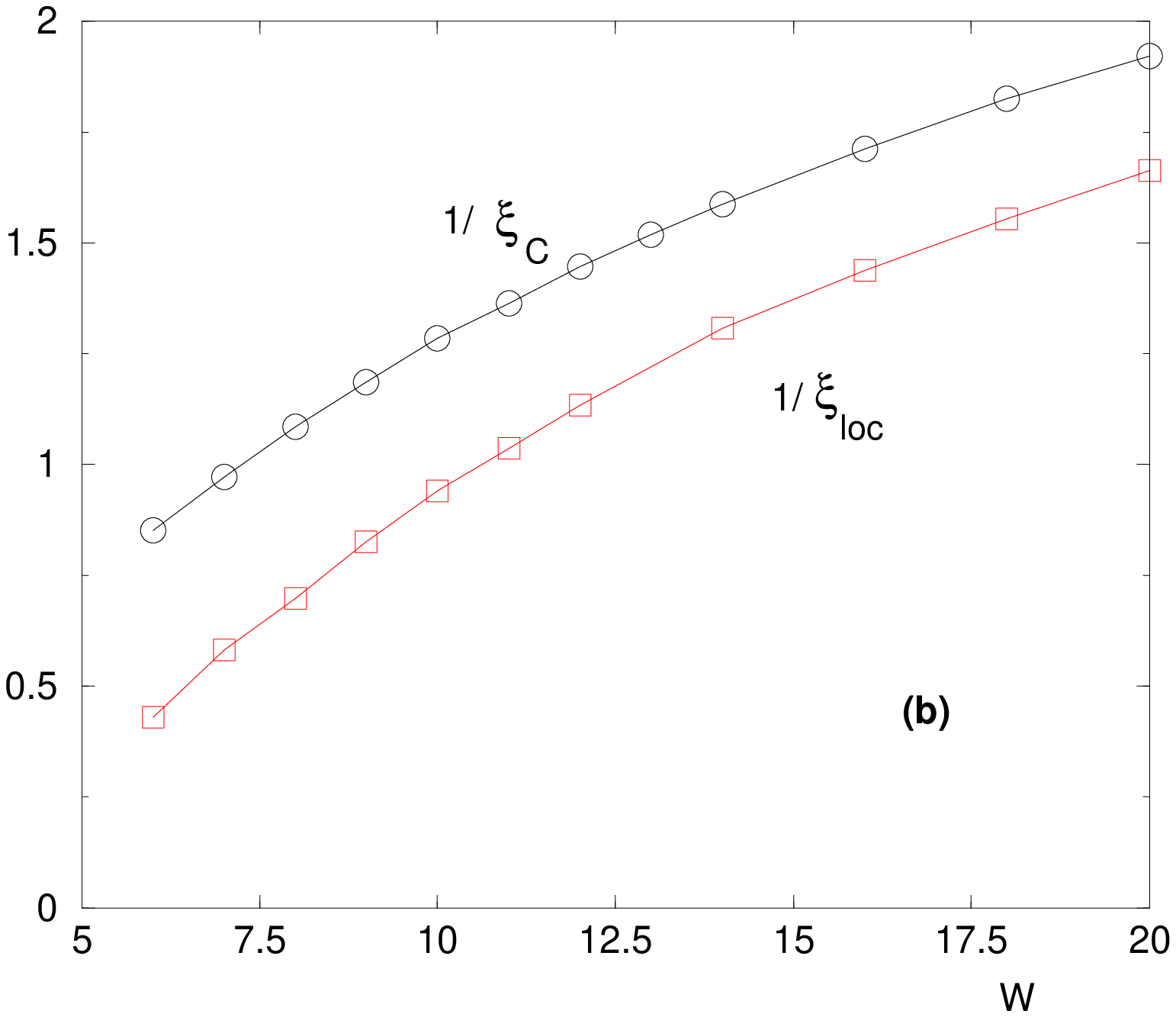}
\caption{  (Color on line) Exponential decay of the typical
real space two-point correlation function $C^{typ}(L)
\equiv e^{\overline{ \ln C(L) }}$ (see Eq. \ref{defcorre} ) in the localized phase :
(a)  Linear decay of $\overline{ \ln V_L }$ 
as a function of $L$  (see Eq. \ref{transloc}).
(b)  Behavior of the slope $1/\xi_{C}(W)$ (inverse of the localization length $\xi_{C}(W)$ of Eq. \ref{correloc}) as a function of the disorder strength $W$ (circles),
 as compared to $1/\xi_{loc}(W)$ (squares)
found previously for the renormalized hopping in configuration space
(see  Eq. \ref{transloc} and Fig. \ref{figtransloc}).
  }
\label{figcorre}
\end{figure}

As recalled in the introduction, the idea that many-body localization
actually occurs in configuration space is very useful, and in this paper,
we have adopted this point of view : we have focused 
on the renormalized hopping
between two configurations separated by a given distance in Fock space,
with the hope that the signatures of the transition would be clearer
for this observable. Nevertheles, it is of course very important to understand
what are the consequences of this localization occuring in configuration space 
for real-space properties. In this section we thus present 
direct calculations of the simplest two-point correlation function
\begin{eqnarray}
C(L)  \equiv \vert < \psi_{mid} \vert c_1^{\dagger} c_L \vert \psi_{mid} >\vert^2
\label{defcorre}
\end{eqnarray}
where $\vert \psi_{mid} >$ is the eigenstate (obtained via exact diagonalization)
of the Hamiltonian of Eq. \ref{hamilton} with free ends at $i=1$ and $i=L$ 
(no periodic boundary conditions, so that these two points are at distance $L$
in real space), whose eigenvalue $E_{mid}$
is in the middle of the many-body energy levels
(this energy $E_{mid}$ fluctuates from sample to sample but remains close
to the central value $E=0$ chosen in (c1) of section \ref{sec-model}).

We show on Fig. \ref{figcorre} (a) the decay with the distance $L$ 
of the disorder-average $\overline{ \ln C(L)}$
for various disorder strengths in the range $6 \leq W \leq 20$ : 
these curves correspond
to an exponential decay with $L$ of the typical value 
$C_L^{typ} \equiv e^{\overline{ \ln C(L) }}$
\begin{eqnarray}
\ln (C_L^{typ}) \equiv \overline{ \ln C_L(W>W_c) }\
 \opsimeq_{L \to \infty} - \frac{L}{\xi_{C}(W)}
\label{correloc}
\end{eqnarray}
where $\xi_{C}$ represents the correlation length that diverges
at the delocalization transition
\begin{eqnarray}
\xi_{C}(W) \opsimeq_{W \to W_c^+} (cst) (W-W_c)^{-\nu_{C}}
\label{xicorreloc}
\end{eqnarray}
We show on Fig. \ref{figcorre} (b) our numerical result
for the slope  $1/\xi_{C}(W)$ as a function of the disorder strength $W$
in the region $6 \leq W \leq 20$, as compared to $1/\xi_{loc}(W)$ 
found previously for the renormalized hopping in configuration space
(see  Eq. \ref{transloc} and Fig. \ref{figtransloc}).
Our conclusion is that up to a numerical prefactor, these two correlation lengths
seen either in the renormalized hoppings in configuration space, or in 
the two-point correlation function in real space, 
contain essentially the same information.
In particular, a three-parameter fit of
 the form of Eq. \ref{xicorreloc} yields 
values for the critical point $W_c$ and for the 
 critical exponent $\nu_C \sim \nu_{loc}$ 
that are compatible with
 the values estimated previously from the data in configuration space.

In the delocalized phase however, it is not clear to us what
 are the theoretical expectations for the decay in $L$ of the two-point correlation function of Eq. \ref{defcorre}, and our numerical data are not sufficiently clear by themselves to
indicate which procedure should be used to fit the data 
in order to obtain information on the critical
behavior in the delocalized phase. 
Further work is needed to clarify this point,
or to find other real-space observables that display a clearer behavior in the delocalized phase.

To summarize this section, our numerical data 
concerning the real-space two-point correlation function of Eq. \ref{defcorre}
indicate that the correlation length measured previously 
in configuration space is essentially equivalent
 to the correlation length measured in real space.
In particular, this shows that the results obtained in configuration space
do not depend to much on the particular choice (c2) of alternate configurations
made in section \ref{sec-model}.

\section{ Conclusion }

\label{secconclusion}

In this paper, we have proposed to study many-body localization transition
via an exact renormalization procedure in configuration space 
that generalizes the
Aoki real-space RG procedure for Anderson localization one-particle models.
For the one-dimensional lattice model of interacting  fermions with disorder studied previously by 
 Oganesyan and Huse \cite{huse}, 
we have studied numerically the statistical properties of the renormalized hopping $V_L$
between two configurations separated by a distance $L$ in configuration space.
Our numerical results are compatible with
 the existence of a many-body localization transition
 at a finite disorder strength of order $W_c \sim 5.6$.
In the localized phase $W>W_c$, we have found that the typical 
renormalized hopping $V_L^{typ} \equiv e^{\overline{ \ln V_L }}$
decays exponentially in $L$ as  $ (\ln V_L^{typ}) \simeq - \frac{L}{\xi_{loc}}$
 and that the localization length diverges as $\xi_{loc} \sim (W-W_c)^{-\nu_{loc}}$ with the critical exponent of order $\nu_{loc} \simeq 0.45$. 
In the delocalized phase  $W<W_c$, we have found that the renormalized hopping
$V_L$ remains a finite random variable $V_{\infty}$ as $L \to \infty$, 
and that the typical asymptotic value $V_{\infty}^{typ} \equiv e^{\overline{ \ln V_{\infty} }}$ presents an
essential singularity $(\ln V_{\infty}^{typ}) \sim - (W_c-W)^{-\kappa}$ with an exponent of order $\kappa \sim 1.4$.
We have argued that the analogy with
Anderson localization on the Cayley tree is important as an example
of Anderson transition on a space of infinite dimension
(in the sense of Eq. \ref{expo}) that presents essential singularities
and that it suggests a specific form of finite size scaling
that we have tested.
Even if the numerical values of the exponents are not expected to be precise, 
as a consequence of the limited system sizes studied $L \leq 14$, we hope
that the scaling laws that emerge are valid. Of course, it would be 
very useful
in the future to test these results with other numerical methods
 like the Density Matrix Renormalization Group method,
that allow to study these interacting one-dimensional models for
 much bigger system sizes \cite{schmitteckert,molina,MacK,znidaric}.
 Finally, we have shown that the present analysis in configuration space
 is compatible with the localization properties displayed by
the simplest two-point correlation function in real space.

In conclusion, the reformulation of many-body localization problems
as Anderson localization models in configuration space 
raises the question of Anderson localization on specific
 networks (see \cite{Berkovits_complex} and references therein), 
that are completely different from the regular lattices
that have been considered in the field of one-particle models.
For a many-body problem defined on a domain of size $L^d$,
the number of configurations (i.e. the nodes of the network)
grows exponentially ${\cal N}_L \propto e^{(cst)L^d}$.
Each configuration has a different connectivity
 in this space of configurations, but it is 
typically of order $L^d$ (assuming a finite density of fermions,
with a finite number of short-range hopping for each fermion).
Besides the interest in specific many-body models,
an important issue is of course to understand which properties
of this complex network are relevant to determine the universality class
of the corresponding Anderson transition.

\section{Acknowledgements}

It is a pleasure to thank D.A. Huse, A.D. Mirlin, T. Prosen and M. Znidaric
for useful discussions and correspondence.
After this work was completed, we became aware of the work \cite{pal_huse}
that suggests an infinitite-randomness scaling for the
many-body localization transition of a quantum spin chain.

\end{document}